\documentclass{aa}

\usepackage{graphics}
\usepackage{times}

\begin{document}

\thesaurus{06.                  
           (06.19.3;            
            06.13.1;            
            02.13.2;            
            03.13.1             
            )}

\title{An approximate self-consistent theory of the
magnetic field 
of fluted penumbrae
}

\author{T.\ Neukirch\inst{1} \and P.\ C.\ H.\ Martens\inst{2}}

\institute{School of Mathematical and Computational Sciences,
University of St.\ Andrews,
St.\ Andrews,
KY16 9SS,
Scotland\\
email: thomas@dcs.st-and.ac.uk 
\and
Space Science Department of ESA at Goddard Space Flight Center,
Code 682.3,
NASA/GSFC,
Greenbelt, MD 20771\\
email: pmartens@esa.nascom.nasa.gov}

\offprints{T.\ Neukirch}

\date{Received date; accepted date}
\titlerunning{Self-consistent theory of fluted penumbrae}
\authorrunning{Neukirch \& Martens}

\maketitle

\begin{abstract}
A self-consistent mathematical description of 
the magnetic field of
fluted sunspot penumbrae is
presented. This description is based on an expansion of the nonlinear
force-free magnetohydrostatic equations written in cylindrical coordinates.
The lowest order solutions are mathematically 
equivalent  
to laminated
force-free equilibria in Cartesian geometry.  
The lowest order
solutions have no toroidal component of the magnetic field and the
magnetic pressure does not vary with azimuth but the solutions allow
arbitrary variations of the magnetic field components with
azimuth. Explicit solutions are presented
which have a realistic radial profile of the magnetic field strength
and reproduce the basic features of the observations.

\keywords{Sunspots -- Sun: magnetic fields -- MHD -- Methods: analytical}
\end{abstract}

\section{Introduction}

Sunspot penumbrae show structures in the form of alternating 
dark and bright fibrils extending in the radial direction across the penumbra.
Observations by Beckers \& Schr\"o\-ter (\cite{beckers:schroeter69}) have already
suggested that the magnetic field inside the dark filaments might be close to
horizontal whereas the field in the bright filaments is less inclined with
respect to the vertical. This magnetic field structure has implications for
the Evershed flow which was also suggested to be associated mainly with
the dark filaments. The association of the Evershed flow with
the dark penumbral fibrils has recently 
been confirmed by Shine et al. (\cite{shine:etal94}) and
Rimmele (\cite{rimmele94,rimmele95b}). In these observations, it was also found that the Evershed effect is a
quasi-periodic time-dependent phenomenon 
which
extends beyond the penumbral boundary (Rimmele \cite{rimmele95b}). 
Furthermore, there
is evidence (Rimmele \cite{rimmele95a}) that 
the Evershed flow occurs in thin channels which
are elevated above the continuum height for most parts of the penumbra.
Very recently, Westendorp-Plaza et al. (\cite{westendorp-plaza:etal97}) have
actually for the first time identified the regions of return flux for the
field lines carrying the Evershed flow. Their observations support the view that
the Evershed flow is carried by low-lying field lines.

The suggested spatial variation of the inclination
angle of the magnetic field with azimuth was also found by 
Lites et al. (\cite{lites:etal90}) and 
Degenhardt \& Wiehr (\cite{degenhardt:wiehr91}). 
Recently, very high resolution observations with the Lockheed tunable
filtergraph on the Swedish Solar Observatory in La Palma 
(Title et al. \cite{title:etal93}) showed that the magnetic field of a sunspot
penumbra indeed 
varies between nearly horizontal in the dark penumbral filaments
and much less inclined in the bright filaments. 
Similar results where obtained by Rimmele (\cite{rimmele95a})
and StanchfieldII et al. (\cite{stanchfieldII:etal97}).
The observations of Title et al. (\cite{title:etal93}) 
show that the mean inclination of the magnetic field increases
from $40^\circ$-$50^\circ$ to $70^\circ$-$75^\circ$ across the penumbra
and that there is a rapid azimuthal variation of the inclination angle of
about $18^\circ$. Little or no variation of the total field strength
with azimuth is found by Title et al. (\cite{title:etal93}) 
and Rimmele (\cite{rimmele95a}) (see, however, the discussion in StanchfieldII et al. (\cite{stanchfieldII:etal97})).
On the basis of these observations, Title et al. (\cite{title:etal93}) proposed
a tentative model for the magnetic field structure, which resembles
that of laminated force-free fields (Low \cite{low88a,low88b}). Since these
fields only exist for Cartesian and spherical geometry, the model
of Title et al. (\cite{title:etal93}) could so far not be backed up by a self-consistent
calculation. 

In a recent paper, Martens et al. (\cite{martens:etal96}) presented a linear force-free
model for a fluted sunspot. The model is able to explain some of the
properties of fluted sunspots. Due to the high azimuthal wave number of
the fluted force-free component, the scale height of this component
is very small and the flutedness is confined to the chromosphere in this
model. Furthermore, the magnetic loops corresponding to the dark filaments
are very short in this model and the filaments do consist of a series of loops.
This implies that the Evershed flow would have to consist of
a phase-coordinated flow along many short loops. There is no explanation how
this could be achieved. Therefore there is a clear need for improved models.

The fundamental problem posed by the observations
is that a genuinely three-dimensional solution of the magnetohydrostatic
equations is necessary. Though there has been some pro\-gress concerning
analytical solutions in three dimensions recently (e.g. Neukirch \cite{neukirch95,neukirch97}), these
solutions do either not apply to the problem or they suffer from the
same deficiencies as the linear force-free solutions. A different approach
is therefore necessary.

In the present paper we present a method which allows us to
find a self-consistent
version of the model originally put forward by 
Title et al. (\cite{title:etal93}).
The method is based on an asymptotic expansion of the three-dimensional
force-free equations. The lowest order of this expansion has solutions
which are formally equivalent to laminated force-free fields and a large
number of different solutions can be found. These solutions can be
used to model the magnetic field of 
the penumbra of a fluted sunspot.

In the following, we first
briefly review the theory of laminated force-free fields and then
present the expansion procedure. A specific solution which is useful
as a model for a fluted penumbral magnetic field is
presented and its properties are discussed especially in connection
with a theoretical explanation of the Evershed flow.

\section{Basic Theory}
\subsection{Non-existence of laminated force-free fields in cylindrical coordinates}
We assume that the 
magnetic field of the sunspot penumbra is force-free and therefore obeys
the equations
\begin{eqnarray}
\vec{ j}\times\vec{ B} &=& 0        ,     \label{jcrossb} \\
\nabla\times\vec{ B}  &=& \mu_0 \vec{ j} ,\label{ampere}  \\
\nabla\cdot \vec{ B}  &=& 0        .     \label{divb}
\end{eqnarray}
From Eq. (\ref{jcrossb}) we conclude that
\begin{eqnarray}
\nabla\times \vec{ B} &=& \alpha \vec{ B}  ,\label{alphab} \\
\vec{ B}\cdot \nabla\alpha &=& 0      ,     \label{Bdalpha}
\end{eqnarray}
where we have absorbed $\mu_0$ into $\alpha$.

The observations 
(Degenhardt \& Wiehr \cite{degenhardt:wiehr91}; Title et al. \cite{title:etal93}; Rimmele \cite{rimmele95a})
show that simple sunspots may
have a fluted penumbra with rapid variations of the inclination
of the magnetic field with azimuth ($\phi$), however with almost constant
magnetic pressure at constant radius. Since the azimuthal component
of the magnetic field is small compared to the other components and the
sunspot field can be considered as being force free,
also the electric currents seem to
flow mostly in the meridional planes ($\phi=$constant). The appropriate
model to describe such a magnetic field would be given by laminated non-linear
force-free fields (Low \cite{low88a,low88b}). 
However, these fields do only exist in
Cartesian and spherical coordinates; in cylindrical coordinates
appropriate for the description of sunspots such fields do not exist.
This can be most easily seen as follows. The basic assumption
is
\begin{eqnarray}
\vec{ B} &=& (B_r,0,B_z)    ,     \label{blaminated}  \\
\alpha  &=& \alpha(\phi)      .  \label{alphalaminated}
\end{eqnarray}
Then Eq. (\ref{Bdalpha}) is automatically fulfilled. If we take the curl
of Eq. (\ref{ampere}) and multiply the result by $\vec{ e}_\phi$ we get
\begin{equation}
\vec{ e}_\phi\cdot\nabla\times(\nabla\times\vec{ B}) = \alpha^2 B_\phi
+ \frac{1}{r} \frac{\mbox{d}\alpha}{\mbox{d}\phi}\vec{ e}_\phi\cdot
\vec{ e}_\phi\times \vec{ B} = 0 . \label{curl1}
\end{equation}
On the other hand
\begin{equation}
\vec{ e}_\phi\cdot\nabla\times(\nabla\times\vec{ B}) =
-\nabla^2 B_\phi -\frac{2}{r}\frac{\partial B_r}{\partial \phi}
+ \frac{2 B_\phi}{r^2} = -\frac{2}{r}\frac{\partial B_r}{\partial \phi} .
\label{curl2}
\end{equation}
It follows that $B_r$ may not depend on $\phi$ which precludes flutedness 
for force-free fields in
cylindrical coordinates. This means that only rotationally symmetric equilibria
of this type are possible and therefore it is impossible find an exact 
force-free solution
appropriate for fluted sunspots.

\subsection{Asymptotic expansion method}

However, if we are mainly interested in the penumbral structure, we could
try to find an approximate solution with the desired properties, based on
laminated equilibria. We start by writing Eqs.
(\ref{alphab}) and (\ref{Bdalpha}) together with Eq. (\ref{divb}) in cylindrical
coordinates:
\begin{eqnarray}
\frac{1}{r}\frac{\partial B_z}{\partial \phi} -
\frac{\partial B_\phi}{\partial z} &=& \alpha B_r ,  \label{requa} \\
\frac{\partial B_r}{\partial z}-
\frac{\partial B_z}{\partial r} &= & \alpha B_\phi,  \label{phiequa} \\
\frac{1}{r}\frac{\partial }{\partial r}\left(\frac{1}{r} B_\phi \right) -
\frac{1}{r}\frac{\partial B_r}{\partial \phi} &=& \alpha B_z ,\label{zequa} \\
B_r \frac{\partial \alpha}{\partial r}+
B_\phi\frac{1}{r}\frac{\partial \alpha }{\partial \phi}+
B_z\frac{\partial \alpha}{\partial z} &= & 0    ,            \label{alphaequa}\\
\frac{1}{r}\frac{\partial }{\partial r}\left(r B_r \right)+
\frac{1}{r}\frac{\partial B_\phi }{\partial \phi}+
\frac{\partial B_z }{\partial z} & =&   0           .          \label{divbequa}
\end{eqnarray}
We now introduce a new radial coordinate 
\begin{equation}
\varpi := \frac{r^2-a^2}{2a} ,    \label{rdef}
\end{equation}
where $a$ is a yet unspecified radius located somewhere inside the penumbra
(actually $a$ could be treated as a free parameter which can be used later to
minimise the residual force resulting from the approximation).
If the radial extent of the penumbra is $2R_p$, we now expand the equations above
in the parameter 
\begin{equation}
\epsilon = R_p/a ,
\label{epsilon}
\end{equation}
which we assume to be smaller than one (this is a kind of 
large-aspect-ratio expansion with
$1/\epsilon$ being the aspect ratio). This type of expansion procedure
together with the coordinate transformation Eq. (\ref{rdef})
has been introduced by Kiessling (\cite{kiessling95}) 
in the framework of toroidally
confined plasma equilibria.

With $r=a\sqrt{1+2\epsilon \varpi/R_p}$ we get
\begin{eqnarray}
\lefteqn{\frac{1}{a\sqrt{1+2\epsilon \varpi/R_p}}\frac{\partial B_z}{\partial \phi} 
              = \alpha B_r  ,} \label{transformedr} \\
\lefteqn{\frac{\partial B_r}{\partial z}-
\sqrt{1+2\epsilon \varpi/R_p}\frac{\partial B_z}{\partial \varpi} =  
\alpha B_\phi                    ,  }                 \label{transformedphi} \\
\lefteqn{\frac{\partial }{\partial \varpi}\left(\sqrt{1+2\epsilon \varpi/R_p} B_\phi \right) -}  \nonumber \\
& &\qquad\qquad\frac{1}{a\sqrt{1+2\epsilon \varpi/R_p}}\frac{\partial B_r}{\partial \phi} 
= \alpha B_z              ,                        \label{transformedz} \\
\lefteqn{\sqrt{1+2\epsilon \varpi/R_p}B_r \frac{\partial \alpha}{\partial \varpi}+}\nonumber \\
& & \qquad\qquad B_\phi\frac{1}{a\sqrt{1+2\epsilon \varpi/R_p}}\frac{\partial \alpha }{\partial \phi}+
B_z\frac{\partial \alpha}{\partial z} = 0    ,             \label{transformedalpha}\\
\lefteqn{\frac{\partial }{\partial \varpi}\left(\sqrt{1+2\epsilon \varpi/R_p} B_r \right)+ } \nonumber \\
& & \qquad\qquad \frac{1}{a\sqrt{1+2\epsilon \varpi/R_p}}\frac{\partial B_\phi }{\partial \phi}+
\frac{\partial B_z }{\partial z} =   0      ,               \label{transformeddivb}
\end{eqnarray}
We now introduce the following expansion scheme in $\epsilon$
\begin{eqnarray}
B_r &=& \sum_{n=0}^{n=\infty} \epsilon^n B_r^{(n)}(\varpi,\phi,z)      ,  \label{expansionbr}    \\
B_\phi &=& \sum_{n=1}^{n=\infty} \epsilon^n B_\phi^{(n)}(\varpi,\phi,z)  , \label{expansionbphi}  \\
B_z &=& \sum_{n=0}^{n=\infty} \epsilon^n B_z^{(n)}(\varpi,\phi,z)   ,     \label{expansionbz}    \\
\alpha &=& \sum_{n=0}^{n=\infty} \epsilon^n \alpha^{(n)}(\varpi,\phi,z)  . \label{expansionalpha}
\end{eqnarray}
Note that $B_\phi$ is of order $\epsilon$ whereas all other quantities are of order
$\epsilon^0=1$.
To lowest order in $\epsilon$ we get:
\begin{eqnarray}
\frac{1}{a}\frac{\partial B_z^{(0)}}{\partial \phi}
 &=& \alpha^{(0)} B_r^{(0)}  , \label{zeroorderr} \\
\frac{\partial B_r^{(0)}}{\partial z}-
\frac{\partial B_z^{(0)}}{\partial \varpi} &= &
0                     ,                                                \label{zeroorderphi} \\
-\frac{1}{a}\frac{\partial B_r^{(0)}}{\partial \phi}
&=& \alpha^{(0)} B_z^{(0)} ,                                           \label{zeroorderz} \\
B_r^{(0)} \frac{\partial \alpha^{(0)}}{\partial \varpi}+
B_z^{(0)}\frac{\partial \alpha^{(0)}}{\partial z} &= & 0   ,           \label{zeroorderalpha}\\
\frac{\partial B_r^{(0)}}{\partial \varpi}  +
\frac{\partial B_z^{(0)} }{\partial z} & =&   0  .                     \label{zeroorderdivb}
\end{eqnarray}
Eqs. (\ref{zeroorderr}) - (\ref{zeroorderdivb}) are completely
equivalent to the equations for laminated force-free fields in Cartesian geometry
(Low \cite{low88a}) with the replacements $x:=\varpi$ and $y:=\phi$.

Multiplying Eq. (\ref{zeroorderr}) by $B_z^{(0)}$ and Eq. (\ref{zeroorderz}) by
$-B_r^{(0)}$ and adding the two equations, we get:
\begin{equation}
\frac{\partial}{\partial \phi}\left( {B_r^{(0)}}^2 + {B_z^{(0)}}^2 \right) = 0 .
\label{bsquaredzero}
\end{equation}
Eq. (\ref{zeroorderalpha}) can be solved by
\begin{equation}
\alpha^{(0)} = \alpha^{(0)}(\phi) .
\label{alphaofphi}
\end{equation}
Eq. (\ref{zeroorderdivb}) is solved by introducing a flux function $A(\varpi, \phi, z)$:
\begin{equation}
\vec{ B}^{(0)} = \left( - \frac{\partial A}{\partial z}, 0,
                         \frac{\partial A}{\partial \varpi}\right) .
\label{fluxfunction}
\end{equation}
Inserting Eq. (\ref{fluxfunction}) into Eq. (\ref{zeroorderphi}) we obtain
\begin{equation}
\frac{\partial^2 A}{\partial \varpi^2} +
\frac{\partial^2 A}{\partial z^2}  = 0 .
\label{laplaceequation}
\end{equation}
Eqs. (\ref{laplaceequation}) and (\ref{bsquaredzero})
allow solutions of the form (Low \cite{low88a})
\begin{equation}
A(\varpi,\phi,z) = \mbox{Re}\; \left(G(\xi)\exp[\mbox{i} \gamma(\phi)]\right) ,
\label{generalsolution}
\end{equation}
where $\xi = \varpi + \mbox{i} z$.
So in this approximation we have a field without a $B_\phi$ component that
is potential in the $\varpi$-$z$-plane with an arbitrary modulation in the
$\phi$-direction.

\section{Modeling the penumbra of a fluted sunspot}

\subsection{The Beckers-Schr\"oter profile}

The solution (\ref{generalsolution}) at first sight leaves a large degree
of freedom to model fluted penumbrae. However, if one wants to calculate
fields which reproduce the basic features of the observations the solution
class is considerably restricted. We will illustrate this by considering
a specific example.

We assume that the basic data we have are the absolute value of the
magnetic field ($B(r)$) and the inclination angle $\delta(r,\phi,0)$ 
of the magnetic field with the vertical
at the level $z=0$. Within the framework of the
theory developed so far, we can only represent spots which have little
or no variation of field strength with azimuth. For reasons of mathematical
convenience, we assume that
\begin{equation}
B(r) = \frac{B_0}{1 + (r/L)^2} .
\label{beckersb}
\end{equation}
This is the radial dependence of the field strength deduced by
Beckers \& Schr\"oter (\cite{beckers:schroeter69}) 
from their observations. Other radial profiles
like that of the Schatzman field (Schatzman \cite{schatzman65}) could also be 
dealt with, but the Beckers-Schr\"oter profile has the advantage of 
greater mathematical
simplicty.

From Eq. (\ref{generalsolution}) the magnetic field is given by
\begin{eqnarray}
B_r &=& \mbox{Im}\left(\frac{\mbox{d} G}{\mbox{d}\xi}
                              \exp[i\gamma(\phi)]\right) ,\\
B_z &=& \mbox{Re}\left(\frac{\mbox{d} G}{\mbox{d}\xi}
                              \exp[i\gamma(\phi)]\right) .
\label{Bcomplex}
\end{eqnarray}
The field amplitude is the given by
\begin{equation}
B = \left\vert \frac{\mbox{d} G}{\mbox{d}\xi} \right\vert .
\label{Babsolute}
\end{equation}
With the coordinate transformation (\ref{rdef}) we rewrite the
Beckers-Schr\"oter profile (\ref{beckersb}) as
\begin{equation}
B(\varpi) = \frac{B_0}{1+\displaystyle{\frac{2 a \varpi + a^2}{L^2}}} .
\label{beckersbinomega}
\end{equation}
Modulo a phase factor this profile corresponds to the absolute value
of the complex function
\begin{equation}
F_{BS}:=\frac{\mbox{d} G}{\mbox{d}\xi} =
       \frac{B_0}{\displaystyle{\frac{2 a}{L^2}}\xi + 1 + 
                  \displaystyle{\frac{a^2}{L^2}}}
\label{GBS}
\end{equation}
on the real axis ($z=0$).
The most general complex function $\psi(\xi)$ with the same absolute value
on the real axis can be written as
\begin{equation}
\psi(\xi) = F_{BS}(\xi)\cdot Q(\xi) ,
\label{generalpsi}
\end{equation}
with
\begin{equation}
\vert Q(\xi) \vert = 1 
\label{qone}
\end{equation}
on the real axis ($z=0$). As we will see below the inclusion of an
appropriate function $Q$ is indispensible for a proper representation
of the penumbral magnetic field.
A simple example of such a function $Q$ is given by
\begin{equation}
Q(\xi) = \exp(\mbox{i}q\xi) ,
\label{Qexp}
\end{equation}
with $q$ a real number.
This example also illustrates that it is most convenient to write $Q(\xi)$
in the form
\begin{equation}
Q(\xi) = \exp(\mbox{i} f(\xi)) .
\label{expf}
\end{equation}
The magnetic field components are the given by the expressions
\begin{eqnarray}
\lefteqn{B_r = \frac{B_0 L^2}{2 a}
          \frac{\exp(-\mbox{Im}(f))}
        {\left(\displaystyle{\frac{L^2+a^2}{2a}} +\varpi\right)^2 +z^2}\cdot  }
\nonumber \\
  & &   \qquad      \Big[ \left(\varpi+\frac{L^2+a^2}{2a}\right)
                 \sin(\gamma(\phi)+\mbox{Re}(f)) - \nonumber \\
  & & \qquad \qquad   \qquad \qquad \qquad \qquad
        z\cos(\gamma(\phi)+\mbox{Re}(f))\Big],
\label{Brexplicit}
\\
\lefteqn{B_z =  \frac{B_0 L^2}{2 a}
          \frac{\exp(-\mbox{Im}(f))}
        {\left(\displaystyle{\frac{L^2+a^2}{2a}} +\varpi\right)^2 +z^2}\cdot }
\nonumber \\
  & & \qquad       \Big[ \left(\varpi+\frac{L^2+a^2}{2a}\right)
                 \cos(\gamma(\phi)+\mbox{Re}(f)) + \nonumber\\
  & & \qquad \qquad    \qquad \qquad \qquad \qquad
         z\sin(\gamma(\phi)+\mbox{Re}(f))\Big] .
\label{Bzexplicit}
\end{eqnarray}

Another important piece of information given by the observations is
the inclination angle $\delta$ of the magnetic field vector,
i.e.\ the angle
between the magnetic field vector and the vertical direction.
For the magnetic field with the components given by Eqs. (\ref{Brexplicit})
and (\ref{Bzexplicit}),
the inclination angle is given by the equation
\begin{equation}
\delta(r,\phi,z) = \gamma(\phi) + \mbox{Re}(f)
                    -\arctan\left(\frac{z}{\varpi +
                               \displaystyle{\frac{L^2+a^2}{2a}}}\right) ,
\label{incgeneral}
\end{equation}
where the last term is the phase of $F_{BS}$.
We remark that with the present theory, we can only
represent fields that have an inclination angle which is additive in the
dependence on the azimuth $\phi$ and the radial coordinate $\varpi$ for
$z=0$. Therefore, we can write the inclination as
\begin{equation}
\delta(\varpi,\phi,z)=\delta_{av}(\varpi,z) + \delta_{mod}(\phi) ,
\label{incspecial}
\end{equation}
with the identities
\begin{eqnarray}
\delta_{av}(\varpi,z)&=&\mbox{Re}(f)
                    -\arctan\left(\frac{z}{\varpi +
                              \displaystyle{\frac{L^2+a^2}{2a}}}\right) ,
\label{deltaa}
\\
\delta_{mod}(\phi)&=&\gamma(\phi) .
\label{deltam}
\end{eqnarray}
The task now is to find a complex function $f(\xi)$ which has the
property $\mbox{Im}(f)=0$ for $z=0$ and
which represents the radial variation of the average inclination for
$z=0$. At first sight this might seem easy because we have already presented
such a function above ($f(\xi)=q \xi$). A closer inspection of the
inclination angle and the magnetic field components for values of $z \ne 0$
reveals, however, that a naive choice of $f$ will cause problems. For
$z =0$ the phase of $F_{BS}$ vanishes, but for $z\ne 0$ it contributes
to the inclination angle. Actually for $z \to \infty$ this contribution
tends to $-\pi/2$. Since the average inclination angle should be positive for
$z=0$ this implies that the radial magnetic field component will
change its sign
as $z$ increases unless $\mbox{Re}(f)$  also increases with $z$
sufficiently fast. In other words, the field lines would bend backwards
toward the sunspot center at a finite height $z$. This is
of course not an acceptable solution. The real part of the linear function
$f(\xi)=q \xi$ 
for example does not increase with $z$ and therefore the linear
function is not a viable choice.

On the other hand, if $\mbox{Re}(f)$ increases without bound 
for $z \to \infty$
the sign of
both $B_z$ and $B_r$ will change an infinite number of times as the argument
of sine, respectively cosine tends to infinity. This as well is not 
an acceptable solution.
We therefore have to find a function $f(\xi)$ which
has a real part that tends to a constant as $z$ goes to infinity and fulfills
all the conditions mentioned previously.

A function which fulfills all these conditions is
\begin{equation}
f(\xi) = q \frac{\xi-L_a}{\xi-L_b}
\label{fright}
\end{equation}
where $q$, $L_a$ and $L_b$ are real numbers. The real and imaginary parts of
this function are
\begin{eqnarray}
\mbox{Re}(f) &=& q \frac{(\varpi-L_a)(\varpi-L_b)+z^2}{(\varpi-L_b)^2+z^2} ,
\label{freal}
\\
\mbox{Im}(f) &=& q \frac{(L_a-L_b) z}{(\varpi-L_b)^2+z^2} .
\label{fim}
\end{eqnarray}
For $z=0$ the imaginary part vanishes as required and for $\varpi$ fixed and
$z \to \infty$ the real part has the limit $q$. Other functions may
fulfill these requirements as well, but this form of $f$ has the advantage
of being relatively simple. Of course, if the radial variation of the
inclination angle would be given by observations, then the observations
would prescribe the real part of $f$ for $z=0$. Here, however, we are 
merely interested to illustrate the method developed and therefore
choose a simple form of $f$. Actually, once one has found one $f$ with
the desired properties, it is possible to add any other complex function $\bar{f}$ which is bounded for $z \to \infty$ and $\varpi$ fixed. An example
of such a function is the linear function $\bar{f}=\bar{q} \xi$. This means that
even with the restrictions discussed above one still has considerable freedom
to model the variation of the average inclination angle with radius across
the penumbra.

\begin{table}
\caption{Parameter values used in the example shown in 
Figs.\ \protect\ref{fig1},
\protect\ref{fig2} and \protect\ref{fig3}}
\begin{tabular}{llc}
\hline
Parameter & Meaning                                         & Value \\
\hline
 $r_o$    & outer penumbral radius                          &  1.0  \\
 $r_i$    & inner penumbral radius                          &  0.5  \\
 $a$      & origin of transformed radial coordinate         & 0.8 \\
 $q$      & amplitude of argument of phase factor & $7\pi/9$ \\
 $\delta_{av,i}$ 
          & average inclination at inner boundary
                                                            & $50^\circ$ \\
 $\delta_{av,o}$ 
          & average inclination at outer boundary 
                                                            & $70^\circ$ \\
 $L_a$    & parameter of function $f$                       & $-0.562$   \\
 $L_b$    & parameter of function $f$                       &  $-1.349$  \\
 $\bar{\delta}_{mod}$ &  amplitude of variation             & $18^\circ$ \\
 $m$      & azimuthal wave number                           & 66 \\
 $B_i$    & field strength at inner boundary      & 1500 G \\
 $B_o$    & field strength at outer boundary      & 1000 G \\
 $B_0$    & amplitude of Beckers-Schr\"oter profile         & $1.8$ \\
 $L$      & length scale of Beckers-Schr\"oter profile      & $1.118$ \\

\hline

\end{tabular}
\label{tab1}
\end{table}

\subsection{An example}

In the present paper we restrict our treatment to the function $f$ given in
Eq. (\ref{fright}). We treat the parameter $q$ as a free parameter which we
choose such that we get a satisfying field line shape. The parameters $L_a$
and $L_b$ are determined by imposing the (average) inclination angles at the
inner and outer boundary of the penumbra. This leads to two simple linear
equations for $L_a$ and $L_b$.
\begin{figure*}
\resizebox{\hsize}{!}{\includegraphics{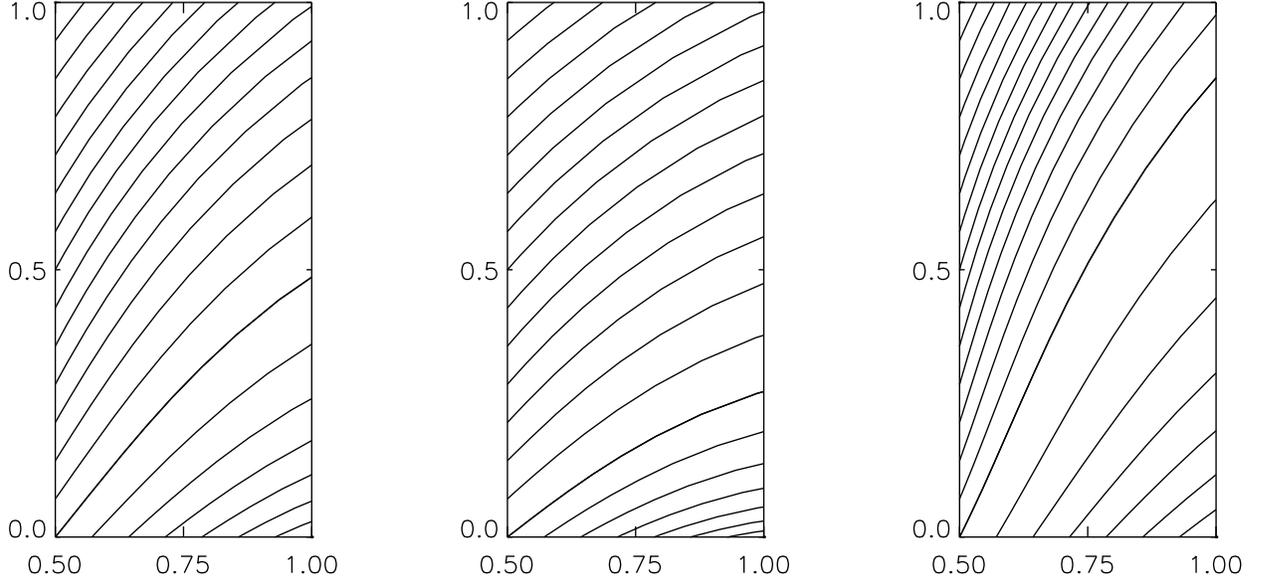}}
\caption{Field line plots in the
         planes $\phi=0$ corresponding to average inclination (left),
         $\phi=\pi/4 m$ corresponding to maximum inclination (middle) and
         $\phi=3\pi/4 m$ corresponding to minimum inclination (right).
         The field lines plotted cross the lower or left boundaries at the same
         location in all three plots to facilitate the comparison. Therefore the
         distance between field lines does not necessarily 
         reflect the strength of the magnetic field. Though not shown here, we emphasize that it is possible to extend the model in such a way that
the return of the low-lying field lines to the surface is included in 
accordance 
with the observations by Westendorp-Plaza et al. 
(\protect\cite{westendorp-plaza:etal97}).}
\label{fig1}
\end{figure*}

For the example we choose the outer boundary of the pen\-um\-bra to be
normalized ($r_o=1.0$)
and the inner boundary of the penumbra to be located at $r_i = 0.5$. The origin
of the transformed coordinate system is choosen as $a=0.8$ locating it
approximately
halfway between the inner and outer boundary of the penumbra. The 
(average) inclination
angles at the inner and outer boundary are chosen as $\delta_{av,i}=50^\circ$
and $\delta_{av,o}=70^\circ$. In this example we have chosen $q=7\pi/9$. The
value of $q$ influences the inclination of the field lines for $z\ne 0$.
Higher values of $q$ give a larger inclination for large
$z$ and vice versa. The resulting values for $L_a$ and $L_b$ are listed 
together with 
the other parameter values in Table \ref{tab1}.
\begin{figure*}
\resizebox{\hsize}{!}{\includegraphics{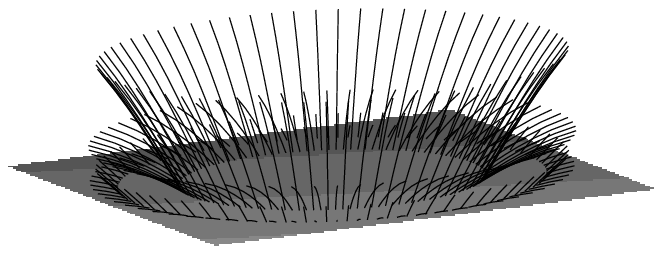}}
\caption{Three-dimensional plot of sets of field lines originating
close to the inner and outer penumbral boundary. The flutedness of the
field is obvious. The azimuthal wave number is $m=66$ in this case.}
\label{fig2}
\end{figure*}

For the modulation of the inclination angle with $\phi$, we choose a simple
harmonic function
\begin{equation}
\delta_{mod}(\phi)=\gamma(\phi)=\bar{\delta}_{mod} \sin( m\phi)
\label{deltamod}
\end{equation}
The parameter $\bar{\delta}_{mod}$ is the amplitude of the inclination
variation which we choose to be $18^\circ$  and
$m$ is the azimuthal wavenumber of the variation which we choose to be
$66$ in accordance with Martens et al. (\cite{martens:etal96}). Of course, any other 
modulation with azimuth would be possible as well as long as it is periodic
with period $2\pi$.

The last two parameters to determine are those of the magnetic field strength,
namely $B_0$ and $L$. These parameters will be determined by 
the measured values
of the magnetic field strength at the inner and outer boundary ($B_i$ and $B_o$). In the present example we have taken $B_i= 1500$ G  and $B_o= 1000$ G. Again, from the two conditions on the inner and outer penumbral
boundary, we get two equations for the parameters $B_0$ and $L$, which can be
easily solved. For the field strengths used we also give the values of $B_0$ and
$L$ in Table \ref{tab1}. The results are shown in Figs.~\ref{fig1} and \ref{fig2}. In Fig.~\ref{fig1}
we show 
field line plots of the field
lines in the planes $\phi=0$, $\phi=\pi/4m$ and
$\phi=3\pi/4  m$ corresponding to average, maximum and minimum inclination.

\begin{figure}
\resizebox{\hsize}{!}{\includegraphics{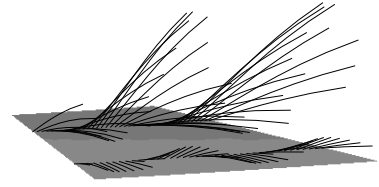}}
\caption{Close-up of a part of Fig.~\protect\ref{fig2}.}
\label{fig3}
\end{figure}
In Fig.~\ref{fig2} we show a set of representative field lines having their
foot points either close to the umbral-penumbral boundary or to the
outer penumbral boundary. The field lines are chosen so that they outline
locations of minimum and maximum inclination. The flutedness of the field is
obvious. In Fig.~\ref{fig3} we show a slice of the full field
over two wavelengths in azimuth to emphasize the
details of the field structure. Especially in Fig.~\ref{fig3} we
see that the field lines with the maximum inclination form long shallow loops
across the penumbra as required for their association with the Evershed effect.

We remark that the model could also be modified to include the return flux observed by Westendorp-Plaza et al. (\cite{westendorp-plaza:etal97}). This
could be done either by choosing different functions $f(\xi)$ (or another
set of parameters with the same function $f(\xi)$) or by extending the
domain beyond the penumbral boundary and shifting the origin of the
transformed radial coordinate system further out.

\section{Discussion}
\subsection{Comparison with the linear force-free model}

There are clear differences between this model and the linear force-free model
of Martens et al. (\cite{martens:etal96}). In the present model the field strength drops off with height more or less like $1/z^2$ for large $z$. The variation 
of the inclination angle with height is 
completely determined by the function $f$ we choose, but in any case it will
have a much weaker dependence on $z$ than the linear force-free model.
In the linear force-free model, the magnetic field drops off exponentially
and the scale height of the fluted part is very small.
This is actually one of the shortcomings of that model.

Another difference is that the parts of the penumbra where the field is nearly
horizontal do not form rather short loops but quite extended radial loops
which can more easily account for the observed Evershed flow.

Whereas in the linear force-free model the direction and the
amplitude of the current density
are determined by the magnetic field because 
$\alpha$ is a constant, 
in the present model $\alpha^{(0)}$ is a function of the azimuth.
The relation between $\alpha^{(0)}$ and $\gamma$ can be calculated either from
Eq.\ (\ref{zeroorderr}) or Eq. (\ref{zeroorderz}). We get
\begin{equation}
\alpha^{(0)}= -\frac{1}{a} \frac{\mbox{d}\gamma}{\mbox{d}\phi}
\label{alphaexplicit}
\end{equation}
With the form of $\gamma$ given in Eq. (\ref{deltamod}), we obtain
\begin{equation}
\alpha^{(0)}= -\frac{m}{a}\bar{\delta}_{mod} \cos(m\phi)
\label{alphacos}
\end{equation}
Since the cosine has its extrema where the sine has its zeros and vice versa,
we have the situation that the maximum current density flows along field lines
having the average inclination, whereas the current density vanishes along
field lines having the maximum or minimum inclination. Furthermore the direction of the current flow changes its sign every half wave length in azimuth. This coincides exactly with the discussion given in Title et al. (\cite{title:etal93}) and sketched
in their Fig.~17. This vindicates our approach because it was our aim
to come up with a self-consistent version of their schematic model.

\subsection{Quality of the approximation}

An important point to investigate is the quality of the approximation
scheme that we have presented. A good way to do this is to investigate
the magnitude of the residual force due to the approximation. To be able
to judge the quality of the approximation, we need to compare the residual
force to a quantity of the same dimension. A convenient measure for
the strength of the force is $B^2(r,z)/\mu_0 L$. We then obtain
\begin{equation}
\mu_0 L\frac{ \vert \vec{ j}\times\vec{ B} \vert}{\vert \vec{ B} \vert^2} =
\frac{\mu_0 L \vert j_\phi\vert}{\vert \vec{ B}\vert} =
\frac{ L \left\vert\displaystyle{\frac{r}{a}}-1\right\vert
        \displaystyle{\frac{\partial B_z}{\partial \varpi}}}{\vert\vec{ B}\vert}.
        \label{joverb}
\end{equation}
Note that the residual force has only a poloidal
component. One can immediately see
that the residual force vanishes at the origin of the transformed
coordinate system ($r=a$).
\begin{figure}
\resizebox{\hsize}{!}{\includegraphics{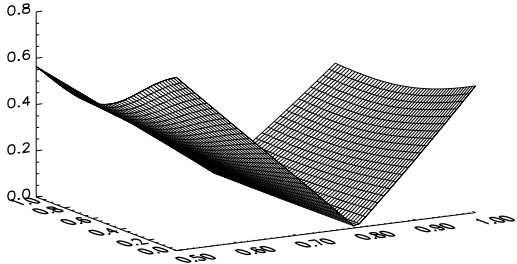}}
\caption{Surface plot of the relative strength of the residual 
$\vec{ j}\times \vec{ B}$-force above the $r$-$z$-plane for $\phi=0$. 
The magnitude of the residual force stays well below $1$, indicating that
the quality of the approximation is reasonable.}
\label{fig4}
\end{figure}
In Fig.~\ref{fig4} we show a surface plot of the residual force as
function of $r$ and $z$ for $\phi=0$. The plots for the locations of
minimum and maximum inclination do not differ very much from this
plot. It can be seen that the residual force is small close to the origin
of the transformed radial coordinate system and then rises almost linearly
in $r$ away from that origin. This can be understood by an inspection of
Eq. (\ref{joverb}). Most of the variation of the residual force expression
obviously comes from the factor $\vert r/a -1\vert$. The factor $\partial B_z/
\partial \varpi / B$ just seems to be a minor modulation of the first factor.
The plot also shows the limits of the approximation scheme. At the boundaries (in $r$) the dimensionless residual force has reached a value of almost $0.6$ which is at the very limit of what can be considered as acceptable for a small parameter. On the other hand, we have only considered the lowest order of the expansion scheme and higher order corrections could make the representation of the field even better. One should also keep in mind that by an expansion procedure like this one can usually not expect to get a convergent series but only 
an asymptotic series (though in a mathematically rigorous sense we would still need to prove that the series is indeed asymptotic).

We mention as a possibility that
in principle, one could integrate the amplitude of the
force over the volume under consideration and minimise this integral
with respect to the parameter $a$. This would give a kind of
optimum value
for the location of the origin of the transformed $r$-coordinate.
Since we expect that the increase in accuracy achieved by such a procedure
will be small we have not carried this out here.

\section{Conclusions}

We have presented an expansion scheme that allows us to calculate self-consistent force-free solutions of the magnetohydrostatic equations
with the basic properties of fluted sunspot penumbrae. The lowest order
of the expansion produces equations which are mathematically equivalent to the
equations describing laminated force-free solutions in Cartesian coordinates.
The magnetic field then has no toroidal component in the lowest order and
the magnetic field strength does not vary with azimuth. The field inclination,
however, can have an arbitrary variation with azimuth. In comparison with
the force-free model by Martens et al. (\cite{martens:etal96}) 
the scale height of
this variation is generally much larger.

We have calculated explicit solutions which have a profile of the magnetic
field strength in the plane $z=0$ that exactly matches the profile given by
Beckers \& Schr\"oter (\cite{beckers:schroeter69}). The field lines emerging from the plane $z=0$
extend across the penumbra in the radial direction and for the maximum
inclination angle stay relatively shallow. 
Our model provides for long, low-lying radial loops that return to the
photosphere just beyond the boundary of the penumbra, as is required for the
Evershed flow, and confirmed by recent observations of Westendopr-Plaza et al.
(\cite{westendorp-plaza:etal97}).
 
\begin{acknowledgements}
The authors thank Aad van Ballegooijen and Pascal D\'emoulin for useful discussions and the referee for helpful comments. TN acknowledges
support by a PPARC Advanced Fellowship.
\end{acknowledgements}

\end{document}